\documentclass[pra,twocolumn,superscriptaddress,aps,showpacs,amsmath,amssymb,nofootinbib]{revtex4}
\usepackage{graphicx}
\usepackage{bm}
\usepackage{amsmath}
\newcommand{\ignore}[1]{}
\newcommand{\bce}{\begin{center}}
\newcommand{\ece}{\end{center}}
\newcommand{\beq}{\begin{equation}}
\newcommand{\eeq}{\end{equation}}
\newcommand{\beqa}{\begin{eqnarray}}
\newcommand{\eeqa}{\end{eqnarray}}

\begin{document}
\title{Beyond standard two-mode dynamics in Bosonic Josephson junctions}

\author{B. Juli\'a-D\'\i az}
\affiliation{Departament d'Estructura i Constituents de la Mat\`{e}ria,\\
Universitat de Barcelona, 08028 Barcelona, Spain}
\affiliation{Institut de Ci\`encies del Cosmos,
Universitat de Barcelona, 08028 Barcelona, Spain}

\author{J. Martorell}
\affiliation{Departament d'Estructura i Constituents de la Mat\`{e}ria,\\
Universitat de Barcelona, 08028 Barcelona, Spain}

\author{M. Mel\'e-Messeguer}
\affiliation{Departament d'Estructura i Constituents de la Mat\`{e}ria,\\
Universitat de Barcelona, 08028 Barcelona, Spain}

\author{A. Polls}
\affiliation{Departament d'Estructura i Constituents de la Mat\`{e}ria,\\
Universitat de Barcelona, 08028 Barcelona, Spain}
\affiliation{Institut de Ci\`encies del Cosmos,
Universitat de Barcelona, 08028 Barcelona, Spain}

\begin{abstract}
We examine the dynamics of a Bose-Einstein condensate 
in a symmetric double-well potential for a broad range 
of non-linear couplings. We demonstrate the existence 
of a region, beyond those of Josephson oscillations 
and self-trapping, which involves the dynamical excitation 
of the third mode of the double-well potential. We 
develop a simple semiclassical model for the coupling 
between the second and third modes that describes 
very satisfactorily the full time-dependent dynamics. 
Experimental conditions are proposed to probe this 
phenomenon.
\end{abstract}


\maketitle

\section{Introduction}

Clouds of cold bosonic atoms exhibit a variety of quantal 
effects on the mesoscopic scale. The initial developments 
dealt with dilute weakly interacting gases, and most of the 
experimental results could be explained by means of the 
Gross-Pitaevskii (GP) 
equation~\cite{Leggett01,pethick2002,pitaevski2003,gatiobert2007}. 
In recent years, the possibility of controlling the atom-atom 
scattering length through Feshbach resonances opened a broad 
range of possibilities for cold atomic clouds, connecting the 
physics of cold atoms to the physics of strongly interacting 
systems; for recent reviews see~\cite{Lewen2007,bloch-rmp}.

Here we consider Bosonic Josephson junctions (BJJ) as our 
benchmark. The GP approach successfully predicted the existence 
of Josephson tunneling phenomena in clouds of bosonic 
atoms confined in a double-well potential~\cite{Smerzi97,Raghavan99}. 
The physics of BJJ can be well understood by further assuming 
a two-mode description of the condensate wave function, a 
simplification which correctly explains the existence of 
self-trapped states in BJJ when the non-linear interaction 
strength is increased~\cite{Smerzi97}. This prediction was 
later confirmed experimentally~\cite{Albiez05}. Very recently, 
most of the semiclassical predictions of this two-mode 
approach, dealing with the Rabi to Josephson transition, 
have been confirmed in an internal Josephson experiment~\cite{tilman10}.

However, it is worth noting that the regime of applicability 
of the quantized two mode approximation can extend further: 
Recent examples are the experiments on BJJ, the production 
of number squeezed states, and a non-linear atom 
interferometer~\cite{esteve08,gross10}. These phenomena are 
beyond GP, as they involve entangled states of the atoms in 
the cloud, but can, however, be explained by a requantization 
of the two-mode equations of GP, the Bose-Hubbard 
model~\cite{Milburn97}. Thus, they are still two-mode physics, 
albeit requantized. 

We scrutinize here the predictions of the GP equation 
when the non-linear interaction term is increased to set 
the condensate beyond the limits of validity of the usual 
two-mode truncation. Our focus is on a specific dynamical 
configuration, the evolution of the BEC when initially 
the majority of the atoms are located in one of the 
wells. By studying the oscillations of the population 
imbalance we find that increasing the non-linear coupling 
the amplitude starts to increase, departing from the usual 
self-trapping behavior. We demonstrate that this dynamics 
can still be explained in terms of a few modes of the 
effective GP potential, but that now it is the variation 
of coupling between the second and third modes that 
successfully explains the GP results.

The manuscript is organized as follows, first in section~\ref{sec1} 
the time dependent GP equation and the usual 
two-mode equations are presented, in section~\ref{sec2} we 
present the results obtained increasing the number 
of atoms, comparing the full time-dependent solutions 
of the GP equations with two-mode predictions. Section~\ref{sec4} 
contains a summary and conclusions.

\section{Theoretical description}
\label{sec1}

For definiteness, we consider a dilute gas of $N$ atoms 
at zero temperature and perform our study in 1D. 
Thus our results should be relevant to cigar-shaped 
quasi-1D experiments. Setting $\hbar=m=1$ the evolution 
is described by the GP equation, 
\begin{equation}
\imath \frac {\partial}{\partial t} \psi(x,t) 
=  -\frac {1}{2} \frac {\partial^2}{\partial x^2} \psi(x,t) 
+  V_{\rm eff}[\psi(x,t)] \psi(x,t)\,.
\label{eq:gp}
\end{equation}
$\psi(x,t)$ is normalized to 1, and the time dependent 
effective potential is 
$V_{\rm eff}[\psi(x,t)]=V(x)+ \lambda_0 N |\psi(x,t)|^2\,.$
The relevant parameter in the GP equation is 
$g_{1D}\equiv  \lambda_0 N$, which sets the importance 
of the non-linear term. In this analysis we consider a 
repulsive interaction, $\lambda_0 > 0$. Different values 
of $N$ and $\lambda_0$ produce exactly the same GP evolution 
provided $\lambda_0 N$ is kept fixed. The confining double-well 
potential, $V(x)$, is generated by $V_{\pm} (x)= (x\pm 2)^2 $ 
connected by $V_p(x)=3 (1-x^2)$ in the interval $\mid x\mid \le 0.5$. 

Eq.~(\ref{eq:gp}) is expected to provide accurate results 
for large enough number of atoms. The recent calculations 
reported in Ref.~\cite{ceder2009} show how the 
exact 1D dynamics of the BJJ approaches the GP predictions as 
the number of atoms is increased~\footnote{The authors of 
Ref.~\cite{ceder2009} consider up to 100 atoms, far away from 
the number of atoms considered in this work and in the experimental set up 
of Ref.~\cite{Albiez05}, $N>1000$.}.

The standard two-mode 
approximation~\cite{Smerzi97,Raghavan99,Ananikian2006} 
is obtained by making the following ansatz for the wave 
function:   
$\Psi(x,t) = \Phi_L(x) \sqrt{N_L(t)} e^{\imath \phi_L(t)}
+ \Phi_R(x) \sqrt{N_R(t)} e^{\imath \phi_R(t)}$. 
The left and 
right modes ($\Phi_{L(R)}=1/\sqrt{2}(\Phi_1\pm \Phi_2)$) 
are constructed from the ground state, $\Phi_1$, and the 
first excited state, $\Phi_2$, of the stationary GP equation. These 
in turn are obtained numerically by an imaginary time 
evolution method. Then a set of equations relating the 
population imbalance, $z(t)=(N_L(t)-N_R(t))/N$ and the 
phase difference, $\delta\phi(t)=\phi_R(t)-\phi_L(t)$ are 
easily derived~\footnote{These are the so-called standard 
two mode equations. Similar conclusions are obtained by 
using the improved two-mode equations of Ref.~\cite{Ananikian2006} 
for the same double-well.}, 
\begin{eqnarray}
\dot{z}(t)&=& - 2{\cal K}\sqrt{1-z^2(t)} \,\sin {\delta\phi(t)}
\label{s2m} \\
\dot{\delta\phi(t)}&=&{ N U  z(t)}
+2{\cal K} {z(t)\over \sqrt{1-z^2(t)}} \, \cos{\delta\phi(t)}\,,\nonumber
\end{eqnarray}
where 
\beqa
U &=&\lambda_0 \int\! dx \,\Phi_{L}^4 \nonumber \\
{\cal K}&=&  -\int d x
\left( (1/2)\partial_x\Phi_{L} \partial_x\Phi_{R}+\Phi_{L}
\,V(x)\,\Phi_{R}\right)
\eeqa
and  integrals involving mixed products of $\Phi_L \Phi_R$ of order larger
than one are neglected. Some authors characterize 
the dynamics with the variable $\Lambda=N U /(2{\cal K})$. 
In our double-well we find, using the modes computed at 
$g_{1D}=0$, ${\cal K}=7.9\times10^{-3}$ and $U/\lambda_0=0.47$, 
which gives $\Lambda=29.7 \lambda_0 N$. The critical value 
of $\Lambda_c$ to have self-trapping within this two mode 
approximation~\cite{Smerzi97}, for ($z(0)=1$, 
$\delta\phi(0)=0$), is $\Lambda_c=2$, which translates 
into a critical value for $g_{1D}^{(1,2)}=0.067$, where the 
superscript $(1,2)$ refers to the states involved in the 
tunneling dynamics.

\begin{figure}[t]
\includegraphics[width=0.98\columnwidth, angle=0, clip=true]{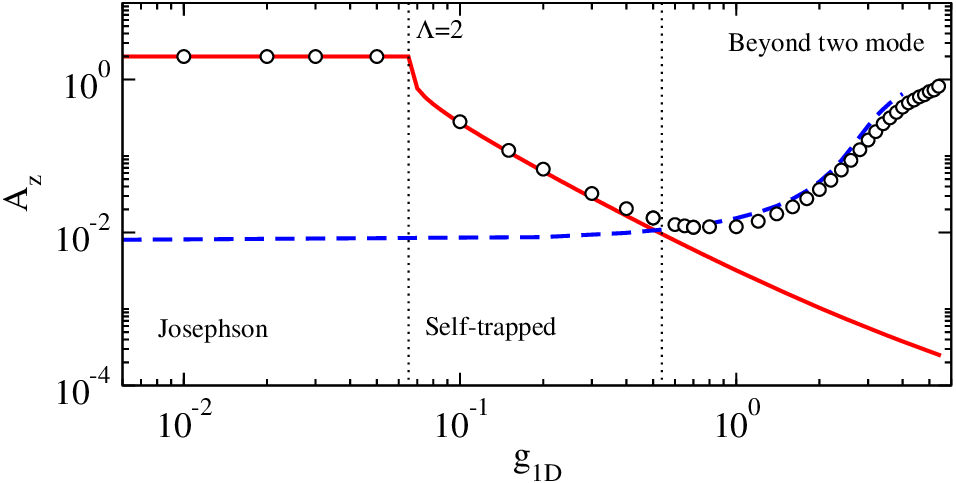}
\caption[]{Maximum amplitude of the imbalance oscillations, 
$z_{\rm  max}-z_{\rm min}$ computed with GP (circles) 
as a function of $g_{1D}$. The solid (red) line 
is the classical two-mode prediction using Eqs.~(\ref{s2m}). 
The dashed (blue) line is a two-mode calculation 
using modes (2,3) as explained in the text.
\label{fig:fig1}}
\end{figure}

\section{Results}
\label{sec2}

\begin{figure}[t]
\includegraphics[width=0.98\columnwidth, angle=0, clip=true]{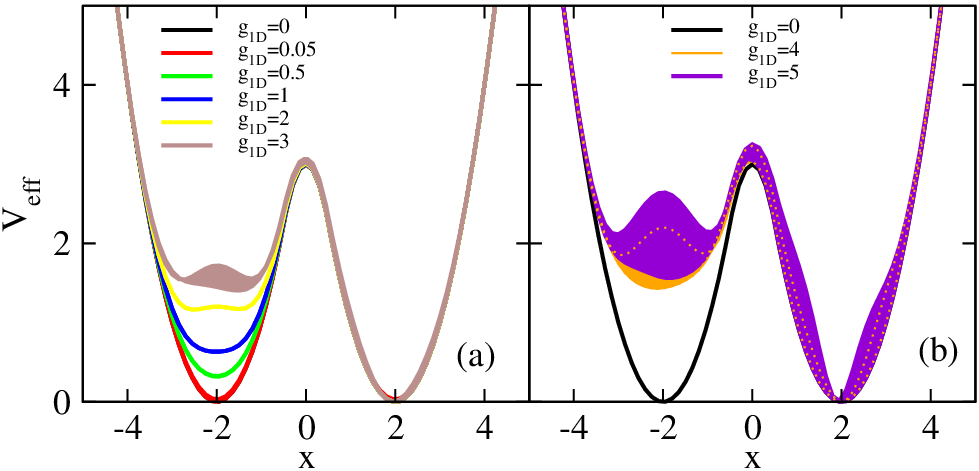}
\caption[]{(color online) 
Effective potential $V_{\rm eff}(x,t)=V(x)+g_{1D} |\Psi(x,t)|^2$ 
for different values of $g_{1D}$. The bands are generated by 
joining $V_{\rm eff}(x,0)$ and $V_{\rm eff}(x,t_{z_{\rm min}})$, 
as explained in the text. The key to the various lines is 
shown in each panel.  
\label{fig:fig2}}
\end{figure}

In all the calculations, except those in 
section~\ref{crit}, we have $\psi(x,t=0)=\Phi_L(x)$ 
(obtained by previously computing $\Phi_1$ and $\Phi_2$ 
of the GP for each $g_{1D}$), corresponding to the case 
of all atoms being on the left well, $z(0)=1$, at $t=0$. 
We study the dynamics for increasing values of $g_{1D} \ge 0 $, 
going from the Rabi regime, $g_{1D}=0$, to the Josephson 
and self-trapped dynamics, and further beyond the range 
of validity of the usual two-mode approximation. 

Results are shown in figure~\ref{fig:fig1}. There 
we compare the numerically determined GP amplitudes of 
$z(t)$, empty circles, with the semiclassical two mode 
prediction of Eq.~(\ref{s2m}), solid (red). At 
$g_{1D} < g_{1D}^{(1,2)}$ there is no self-trapping and $z(t)$ 
oscillates between $+1$ and $-1$, thus leading to a 
constant maximal amplitude, 
$A_z\equiv z_{\rm max}-z_{\rm min}=2$. With increasing 
interaction strength, near $g_{1D}^{(1,2)}$, self-trapping 
appears, the atoms become increasingly confined in the 
left well, and $A_z$ decreases abruptly. In all this range 
the semiclassical model predictions are very successful, 
covering the well known Josephson and self-trapped regimes. 
This range of $g_{1D}$ is the one recently explored 
experimentally in Ref.~\cite{tilman10}. 

With further increase of $g_{1D}$, deviations begin to 
appear. Whereas the semiclassical two-mode model predicts a 
smooth decrease of $A_z$, the GP calculations, empty circles, show 
a smooth reappearance of tunneling between the two wells. This 
is the new phenomenon that we will now discuss and interpret.  

Figure~\ref{fig:fig2} shows the effective potential 
defined in Eq.~(\ref{eq:gp}) for several values of 
$g_{1D}$ at two different times, $t=0$ and $t=t_{z_{\rm min}}$ 
which correspond to the time of the first minimum of 
the population imbalance. In this way the band covers 
the variation of the effective potential during the 
simulation. When $g_{1D}\lesssim g_{1D}^{(1,2)}$ the non-linear 
contribution is fairly small, and 
$V_{\rm eff}(x,t) \simeq V(x)$ at all times. This 
corresponds to the Rabi and Josephson regimes, with 
maximal oscillations of the population. Increasing $g_{1D}$ 
further, $g_{1D}^{(1,2)} \lesssim g_{1D} \lesssim 3$, the value of 
$V_{\rm eff}(x,t)$ in the left well is increased, but leaving 
the value in the right well almost unchanged. This is a 
direct consequence of self-trapping. In this regime, the 
effective potential changes very little with time, see 
fig.~\ref{fig:fig2} (a).  Further increasing $g_{1D}$, 
$3 \lesssim g_{1D} \lesssim 5$, the potential on the 
left well increases, and $V_{\rm eff}$ does begin to change 
appreciably with time. Still, the dynamics remains 
self-trapped but with larger oscillation amplitudes in $z(t)$. 

\begin{widetext}

\begin{figure}[t]
\includegraphics[width=0.7\columnwidth, angle=0, clip=true]{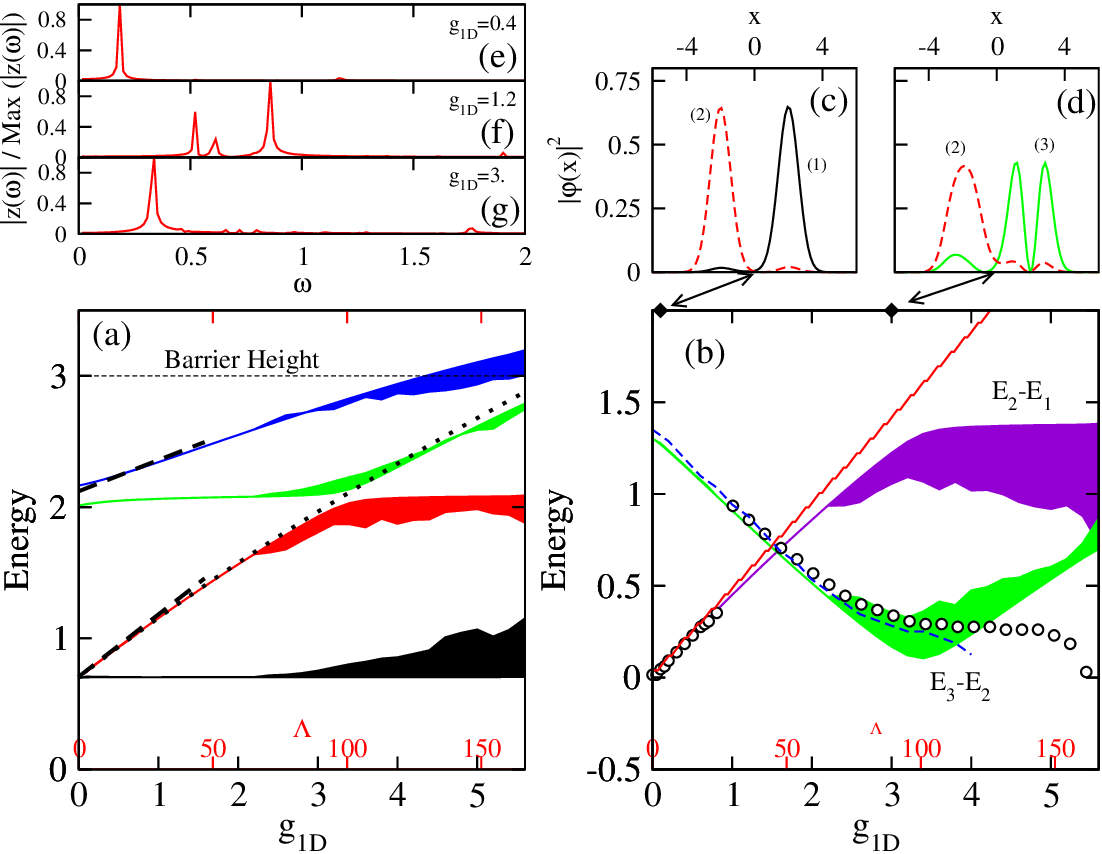}
\caption[]{(a) First four eigenvalues of $V_{\rm eff}$ at 
$t=0$ and $t=t_{z_{\rm min}}$, thus generating the bands. 
The dashed black lines are the first order perturbation 
theory calculation described in the text. The dotted 
line is the average energy of the initial 
state, $\langle \psi(x,0)| H | \psi(x,0)\rangle$. 
(b) Energy differences $E_2-E_1$ (violet band) 
and $E_3-E_2$ (green band) compared to twice the 
frequencies of oscillation of the population 
imbalance computed with GP, empty circles. The solid 
(red) line is the usual two-mode calculation using 
Eqs.~(\ref{s2m}). The dashed (blue) line is a two-mode 
calculation using modes (2,3) as explained in the 
text. (c) and (d) depict the two relevant eigenstates 
of the effective potential (computed at $t=0$) entering 
into the dynamics for $g_{1D}=0.1$ and 3, respectively. 
(e), (f), and (g) depict the normalized Fourier 
transform of $z(t)$, $|z(\omega)|$, for 
$g_{1D}=0.4, 1.4$ and $3$ respectively. 
\label{fig:fig3}}
\end{figure}
\end{widetext}

\subsection{Analysis of the effective potential}

To clarify the discussion and further gauge the smallness 
of the changes in the potential with time, we have 
determined the first four stationary solutions of the 
Schr\"odinger equation built with $V_{\rm eff}(x,t)$, 
with $t$ a fixed parameter. Figure~\ref{fig:fig3} (a), 
shows the eigenvalues thus found numerically. Two values 
of the parameter $t$ have been chosen: $t=0$ and 
$t=t_{z_{\rm min}}$, to form a band for each eigenvalue. 
The band width is seen to remain very small for 
$g_{1D} \lesssim 2.2$, in line with the largely time 
independent $V_{\rm eff}$ depicted in Fig.~\ref{fig:fig2} (a). 
For small interaction strengths, $g_{1D}\lesssim 2$, 
two of the eigenvalues are practically independent of 
$g_{1D}$: they correspond to states 1 and 3, mostly 
located in the right well, which remains unchanged as 
seen in Fig.~\ref{fig:fig2} (a). Their values are thus 
close to the corresponding non-interacting harmonic 
oscillator, $E_{1}^{ho}=\sqrt{2}/2$ and $E_{2}^{ho}=3 \sqrt{2}/2$. 

The other two eigenvalues increase smoothly with the 
interaction strength, they correspond to states 2 and 
4 located mostly in the left well. The increase follows 
the behavior of the effective potential shown in 
figure~\ref{fig:fig2}. The almost linear increase 
in energy of these two eigenvalues can be understood 
treating the non-linear term as a perturbation, then 
\beqa
\delta E_2&=&g_{1D} \langle \,\psi_1^{ho}|\,
|\psi_1^{ho}|^2\,|\psi_1^{ho}\rangle
=g_{1D}/(2^{1/4}\sqrt{\pi})\,,\nonumber \\
\delta E_4&=&g_{1D} \langle \psi_2^{ho}|\, 
|\psi_1^{ho}|^2\,|\psi_2^{ho}\rangle=\delta E_2/2\,,
\label{eq:e24}
\eeqa 
where $\psi_{1,2}^{ho}$ are the first two eigenfunctions 
of the harmonic trap, $V(x)=(x+2)^2$:
\beqa
\psi_{1}^{ho}(x)&=& (\sqrt{2} /\pi)^{1/4}\, e^{-\sqrt{2}
  (x+2)^2\over 2} \,,\nonumber \\
\psi_{2}^{ho}(x)&=&2^{3/4} \, (x+2) \, \psi_{1}^{ho}(x)\,.
\eeqa
These estimates are shown in Fig.~\ref{fig:fig3} (a) 
as dashed lines. \\
\\
As sketched above, in the conventional two mode picture 
one considers only the lowest two stationary solutions 
of the original GP equation,~(\ref{eq:gp}),
and self-trapping occurs due to a misalignment 
of their eigenenergies that suppresses the (Josephson) 
tunneling between the two states. The dynamics thus 
involves two quasi-degenerate states arising from the 
ground states of the unconnected wells.

\subsection{Beyond the usual two-mode dynamics}

What is new here is that near $g_{1D} \simeq 2$ 
$(\Lambda \sim 60)$, the third eigenvalue becomes 
aligned with the second and, since the corresponding 
modes are located in different wells, tunneling of atoms 
is again allowed. But now between modes 2 (left well) 
and 3 (right well). This corresponds to the rise
of $A_z$ in figure~\ref{fig:fig1} beyond $g_{1D} \simeq 1$. 
Figure 3a shows that by increasing the non-linear 
interaction we go from the usual coupling between the 
two lowest modes, 1 and 2, see panel (c), to a coupling 
of the lowest mode of the left-well, 2, to the first 
mode of the right well, 3, see panel (d). This coupling, 
which is zero in absence of interaction, occurs due to 
the large nonlinearity: it deforms the wave functions 
enough to enable the coupling between the ground state 
of one well and the first excitation in the other. It 
is a large and clearly density dependent effect which 
requires a change in the modes used and cannot be 
accommodated by varying the parameters of the usual 
two-mode picture. This effect is different in nature 
from what was reported in Refs.~\cite{tonel2005,carr2007,carr2010} 
where the alignment takes place due to the presence of 
a large enough bias in the system. In our case, it 
is clearly a dynamical phenomenon which happens even 
for symmetric double-well potentials. The role played 
here by the non-linear interaction in modifying the 
single particle states is more similar to the interaction 
blockade effect demonstrated for double-wells with 
few atoms in optical superlattices \cite{cheinet08}.  

Our result is also of different nature to the disappearance 
of self-trapping reported in Ref.~\cite{meier01}. There, 
the authors explore the population of low-energy 
Bogoliubov excitations in the condensates of each of the 
wells finding, using a schematic model, a departure 
from self-trapping due to excitation of such low energy modes. 
By using the time-dependent GP equation to determine 
the condensate wavefunction, the low energy Bogoliubov 
states are already incorporated into its time evolution. 
See Eq. (8.43) and the discussion in section VIII.E in 
Ref.~\cite{Leggett01}. Including them again along the 
lines of Ref.~\cite{meier01} would be redundant.

To further confirm our picture, in panel (b) of 
figure~\ref{fig:fig3} we compare the frequencies of 
oscillation of $z(t)$~\footnote{These frequencies 
correspond to those with the largest amplitudes in 
the direct Fourier transform $|z(\omega)|$ of the 
function $z(t)$, over one Rabi time.} found in the full 
GP calculations, shown as empty circles, to the energy 
differences of the first three modes~\footnote{The 
energy of the fourth mode, see Fig.~\ref{fig:fig3}, 
shows a constant increase, without avoided 
crossings in the range of $g_{1D}$ considered, 
so that this mode remains mostly uncoupled to 
the rest and we will ignore it.}. Up to $g_{1D} \simeq 1$ the 
two mode description with 
states 1 and 2 works very well, but beyond that, the 
appropriate two mode model must involve states 2 and 
3, and tunneling between wells increases again due 
to the progressive alignment between the energies 
of these two modes. The transition from (1,2) to 
(2,3) takes place smoothly, with the region 
$1\lesssim g_{1D}\lesssim 2$ having more than one 
clear peak in the Fourier transform, $z(\omega)$, see 
panel (f) of Fig.~\ref{fig:fig3}, but an extremely 
small oscillation amplitude as seen in Fig.~\ref{fig:fig1}. 
In this region, the dynamics is governed by the 
first three modes coupled pairwise, (1,2) and 
(2,3).

\begin{figure}[t]
\includegraphics[width=0.95\columnwidth, angle=0, clip=true]{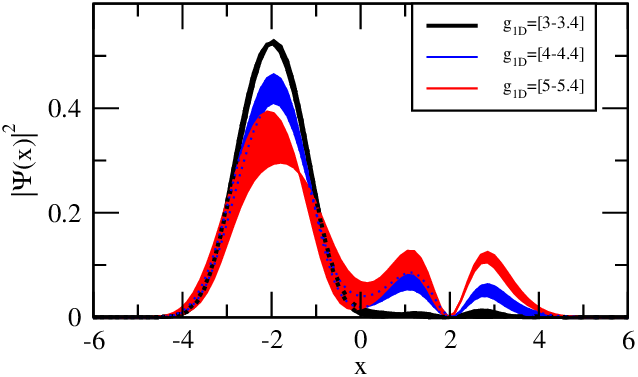}
\caption[]{Density of atoms along the $x$ direction at 
a given time $t=t_{\rm Rabi}/10$ for several values of $g_{1D}$. The 
bands are built by joining the calculations at two different 
values of $g_{1D}$.  
\label{fig:fig4}}
\end{figure}

Above $g_{1D}\simeq 2$, the dynamics is dominated 
by modes 2 and 3, depicted in panel (d) of 
Fig.~\ref{fig:fig3}. The strongest frequency 
extracted from the GP calculation now falls close to 
the $E_3-E_2$ band (green). Following similar arguments
to those in the derivation of Eqs.~(\ref{s2m}) we can 
write down the new two mode equations: We denote 
the second and third modes of $V_{\rm eff}(x,0)$ 
as $\Phi_L$ and $\Phi_R$. A slight generalization of 
Eqs.~(\ref{s2m}) leads to
\begin{eqnarray}
\dot{z}(t)&=& - 2{\cal K} \sqrt{1-z^2(t)} \,\sin {\delta\phi(t)} \\ 
\dot{\delta\phi(t)}&=&\Delta+N\,\bar{U}  z(t)
+2{\cal K} {z(t)\over \sqrt{1-z^2(t)}} \,
\cos{\delta\phi(t)}\, \nonumber 
\label{s2mg} 
\end{eqnarray}
where, 
\beqa
\Delta&=&(E_L^0-E_R^0)+(U_L-U_R)/2\nonumber \\ 
\bar{U}&=&(U_L+U_R)/2 \nonumber \\ 
U_{L(R)} &=&\lambda_0 \int\! dx \,\Phi_{L(R)}^4  \\
E^0_{L(R)}&=&  \int d x
\left( (1/2)\partial_x\Phi_{L(R)} \partial_x\Phi_{L(R)}+\Phi^2_{L(R)}
\,V(x)\right)\,. \nonumber
\eeqa 
Using $g_{1D}=1.2$ to build the modes gives, 
$\Delta=-1.26$, $\bar{U}=0.35$, and ${\cal K}=-0.037$. 
The prediction of this new two-mode model is shown by the 
dashed (blue) lines in Fig.~\ref{fig:fig1} and 
Fig.~\ref{fig:fig3}(b). As can be seen this (2,3) model 
works well in the range $1 \lesssim g_{1D} \lesssim 3.5$, 
giving a good account of both the dominant frequency 
and the amplitude of the imbalance observed in the GP 
simulations. The transition from (1,2) to (2,3) 
coupling reflects also in the appearance of the node 
of the $|\psi(x,t)|^2$ near $x=2$, obtained solving 
the GP equations as seen in Fig.~\ref{fig:fig4}. This is 
an observable feature which should be looked for 
experimentally.

Further increasing $g_{1D}$, $g_{1D}\gtrsim 6$, our 
initial state has an average energy above the 
barrier, see dotted line in Fig.~\ref{fig:fig3} (a), 
thus facilitating the flow of atoms 
between the wells. 
At high enough $g_{1D}$ a certain equilibration of 
the imbalance can be expected, in line with 
Ref.~\cite{ceder2009}, mostly due to the sizeable 
contributions from  modes with energies above the 
barrier. 

\subsection{Critical $g_{1D}^{(2,3)}$ for $z(0) < 1$}
\label{crit}

Until now we have considered only cases with a maximally imbalanced 
initial condition, $z(0)=1$. For other initial conditions, 
$z(0)< 1$ we can estimate the critical value for the 
tunneling between modes (2,3), $g_{1D}^{(2,3)}$,  
using similar arguments to the ones leading to Eq.~(\ref{eq:e24}). 
The value of $g_{1D}^{(2,3)}$ for $z(0)=1$ corresponds to the one 
for which the second energy level (the first of the most 
populated well) equals the third energy level (the second 
of the less populated well). Thus fulfilling $E_1^{ho}+\delta E_2 = E_3$, 
which can easily be solved giving, 
$g_{1D}^{(2,3)}=2^{3/4}\sqrt{\pi}\sim 2.98$, in good agreement with 
Figs.~\ref{fig:fig2} and ~\ref{fig:fig3}. 

Similarly, we can estimate the critical value, $g_{1D}^{(2,3)}$, 
for an initial state with a certain population imbalance 
$z_0\equiv z(0) >0$. In this case, assuming the system 
remains mostly self-trapped, we can define the fraction 
of atoms on the left well, $p_L \equiv (1+z_0)/2$, and on the 
right well, $p_R \equiv (1-z_0)/2$. We can use a similar argument 
as before and assume the wave function of the system is 
essentially a harmonic oscillator wave function at each 
side of the trap. We assume, the left well stays mostly 
on the ground state, while the right side is promoted to 
the first excited one. Treating again the non-linearity 
as a perturbation, we have, 
($E_1\equiv E_{1R}$, $E_2\equiv E_{1L}$, $E_3\equiv E_{2R}$,
$E_4\equiv E_{2L}$),  
\beqa
\delta E_1  &=&  
g_{1D} p_{R} \langle \psi_{1}^{ho} | |\psi_{2}^{ho}|^2 |\psi_{1}^{ho}\rangle
=g_{1D}p_{R}/(2^{5/4}\sqrt{\pi}) \,,\nonumber \\
\delta E_2  &=&  
g_{1D} p_{L} \langle \psi_{1}^{ho} | |\psi_{1}^{ho}|^2 |\psi_{1}^{ho}\rangle
=g_{1D}p_{L}/(2^{1/4}\sqrt{\pi}) \,,
\nonumber \\
\delta E_3  &=&  
g_{1D} p_{R} \langle \psi_{2}^{ho} | |\psi_{2}^{ho}|^2 |\psi_{2}^{ho}\rangle
=(3/2) \delta E_1 \,,
\nonumber \\
\delta E_4  &=&  
g_{1D} p_{L} \langle \psi_{2}^{ho} | |\psi_{1}^{ho}|^2 |\psi_{2}^{ho}\rangle
=(1/2)\delta E_2\,.
\eeqa
The critical condition would correspond to have 
$E_{2}\sim E_{3}$, that is $E_2^{ho}+\delta E_3 = E_1^{ho}+\delta E_2$ 
\beq
{3\over \sqrt{2}} + {3\over 2} {g^{1c}_{1D} p_R\over 
2^{5/4}\sqrt{\pi}}
= 
 {1\over \sqrt{2}} + {g^{1c}_{1D} p_{L}\over 2^{1/4}\sqrt{\pi}} \,,
\eeq
which gives, 
\beq
g^{(2,3)}_{1D}= {4\; 2^{3/4}\sqrt{\pi} \over 4 p_L-3 p_R }\,.
\eeq
For $p_R\ll 1$ we have, 
$
g^{(2,3)}_{1D} \sim 2^{3/4} \sqrt{\pi} \left[1 +  (7/4) p_R\right]\,.
$
Thus, for $z(0)$ close to 1, the critical value, $g_{1D}^{(2,3)}$, grows 
linearly with the initial fraction of atoms in the less populated well. 

\subsection{Possible experimental conditions}
\label{sec3}

The phenomena presented in the previous sections 
rely on the adequateness of the Gross-Pitaevskii 
equation to describe the physics at large enough 
values of $g_{1D}$. The predicted reappearance of 
tunneling of atoms through the barrier beyond 
the usual self-trapped region as we increase $g_{1D}$ 
requires quantum fluctuations to be minimized during 
the whole process. Thus, we need to estimate with 
realistic conditions whether the system can be brought 
to such $g_{1D}$ values while remaining mostly condensed. 

We consider $N$ atoms of $^{87}$Rb trapped inside an axially 
symmetric trap, characterized by $\omega_\perp$, and 
its associated length $a_\perp= (\hbar / M \omega_\perp)^{1/2}$. 
The double-well potential is built with a frequency 
$\omega_x$ inside each well. The scaled 1D coupling, 
$\hat{g}$, can be written in the weakly interacting 
limit~\footnote{As discussed in Ref.~\cite{sala02}, Eq.~(\ref{eq:gp})
can be obtained from the 3D GP one in the weakly interacting limit. 
For the strongly interacting limit the corresponding 1D 
reduction would correspond to 
$\tilde{V}_{\rm eff}=V(x) + \tilde{\lambda}_0 \sqrt{N} |\psi(x,t)|$. 
We have checked that the phenomena described in the paper are 
also present in such reduction.} as~\cite{sala02}, 
\beq
\hat{g}={4 \pi \hbar^2 a_s  \over 2 \pi a_\perp^2  M} = 
{2 \hbar a_s   \omega_\perp  } \,.
\eeq
To obtain meaningful estimates we consider the elongated 
trap conditions of Ref.~\cite{gio08}: 
$\omega_x=2 \pi 44.7$ Hz and $\omega_\perp= 2 \pi 1100$ Hz. 
For these we can estimate the dilution factor, $\eta= n_{1D} a_s$, 
where $a_s$ is the s-wave scattering length and $n_{1D}$ is 
the one-dimensional density. In the Thomas-Fermi limit, this 
estimate is given by 
\beq
\eta = (g_{1D})^{2/3} {\omega_x \over \omega_\perp} {1\over 3^{1/3} 2^{5/3}} \,,
\eeq
and for the trap conditions considered, 
$
\eta \sim   9 \times 10^{-3} (g_{1D})^{2/3}\,.
$
Thus, an elongated trap ensures the diluteness of the 
gas along the direction of the barrier, which is a 
necessary condition for the validity of the Gross-Pitaevskii approximation. 
The diluteness on the longitudinal direction also ensures the non-excitation 
of transverse modes of the elongated trap. This can be checked by 
comparing the transverse chemical potential with the transverse 
quantum, $\hbar \omega_\perp$. The transverse chemical potential 
can be written as, see Eq.~(10) of ~\cite{paul07}, 
$
\mu_\perp = \hbar \omega_\perp ( 1+  2 \eta) 
$
which is well below $2 \hbar \omega_\perp$ for the values of 
$g_{1D}$ considered in this work provided the trap is 
cigar-shaped.

This excitation of the third mode (the first excited 
state of the less-populated well) should show up 
macroscopically 
in the experiments as an increase in the amplitude 
of oscillations of the population imbalance above 
a certain value of the interaction strength. 
Simultaneously, a node should appear in the center 
of the atomic cloud sitting in the less populated 
well, see Fig.~\ref{fig:fig4}. 

Experimentally, the challenge is twofold: first one 
needs to build a double-well potential with a 
single-particle spectrum close to the one described 
here, see Fig.~\ref{fig:fig3}a.
This can be achieved by modifying the parameters of the 
optically produced double-well used in Ref.~\cite{Albiez05}. 
Secondly, one needs to increase the number of atoms to have a
non-linear term able to scan the 
transition described in this article. 
Such an experiment is within 
reach with current experimental setups~\cite{oberp}.

\section{Conclusions}
\label{sec4}

We have demonstrated the possibility of exciting 
higher modes of the double-well potential through 
the dynamics. We have considered an initially 
imbalanced population and have shown 
that for a broad range of interaction energies 
the system remains self-trapped but not due to 
the dynamics of the two lower states of the 
Gross-Pitaevskii equation, as usually accepted, but 
of another mechanism involving a third state. This 
transition from the coupling between the first two 
(1,2) to the next two (2,3) states can be well 
characterized and understood by analyzing the 
static properties of the effective potential 
(including interactions) which due to the fact 
that the system remains mostly self-trapped does 
not vary substantially with time. 

\begin{acknowledgments}

We thank D.W.L. Sprung for useful remarks and a 
careful reading of the manuscript. M. M-M, is 
supported by an FPI grant from the MICINN (Spain). 
B.J-D. is supported by a CPAN CSD 2007-0042 
contract. This work is also supported by Grants 
No. FIS2008-01661, and No. 2009SGR1289 from 
Generalitat de Catalunya.
\end{acknowledgments}


\end{document}